\documentclass[a4paper,10pt,openright,twoside]{article}
\usepackage[english,francais]{babel}
\usepackage[latin1]{inputenc}
\usepackage[T1]{fontenc}

\usepackage[dvips]{graphicx}

\usepackage{url}
\def\auteurTitre{Bernard \textsc{Jacquemin}\up{1}, Aurélien \textsc{Lauf}\up{1}, Céline \textsc{Poudat}\up{2},\\ Martine \textsc{Hurault-Plantet}\up{1} et Nicolas \textsc{Auray}\up{2}}
\def\auteur{B. \textsc{Jacquemin} \textit{et al.}}
 
\def\adresselabo{\up{1}LIMSI CNRS UPR 3251, Orsay (France) \\
		 \up{2}ENST, Paris (France) }
 
\def\courriel{\{Bernard.Jacquemin,Aurelien.Lauf,Martine.Hurault-Plantet\}@limsi.fr \\
              \{Celine.Poudat,Nicolas.Auray\}@enst.fr }

\def\titre{Managing conflicts between users in Wikipedia}

\def\titrecourt{Conflicts in Wikipedia}

\def\piedpage{Social Aspects of the Web 08, Innsbruck, May 6 2008, pp. 87-99.}

\usepackage{fancyhdr}
\title{\titre}
\author{\auteurTitre\\\adresselabo\\\courriel}
\date{}

\begin{document}

\maketitle              

\selectlanguage{english}
\begin{abstract}        
Wikipedia is nowadays a widely used encyclopedia, and one of the most visible sites on the Internet. Its strong principle of collaborative work and free editing sometimes generates disputes due to disagreements between users. In this article we study how the wikipedian community resolves the conflicts and which roles do wikipedian choose in this process. We observed the users behavior both in the article talk pages, and in the Arbitration Committee pages specifically dedicated to serious disputes. We first set up a users typology according to their involvement in conflicts and their publishing and management activity in the encyclopedia. We then used those user types to describe users behavior in contributing to articles that are tagged by the wikipedian community as being in conflict with the official guidelines of Wikipedia, or conversely as being well featured. 
\end{abstract}

\section{Introduction}

The Wikipedia encyclopedia project has become a reference informational resource, and one of most visible sites on the Internet. Amazing and far removed from the Enlightenments spirit -- where the expert and his signature constitute the text quality guarantee --, Wikipedia is based on a very different editorial process.

The whole project is based on a few strong ideological principles, also called \textit{pillars}, \textit{official guidelines} or \textit{fundamental principles} in Wikipedia. First, the goal is clearly to be a generalist encyclopedia project with several linguistic instances that are independently managed. Then, the Wikipedia contents also have to be objective. Wikipedians reckon that the best way to grant the objectivity is to set out a \textit{neutral point of view} (NPOV)\footnote{The articulation between both is performed as follow: "What people believe is a matter of objective fact, and we can present that quite easily from the neutral point of view." (Jimbo Wales, co-founder of Wikipedia, \url{http://en.wikipedia.org/wiki/Wikipedia_talk:Attribution/Role_of_truth}). Thus the Wikipedia's aim at the objectivity is only performed at an \textit{opinion} inventory level, despite their uneven quality on the same page \cite{GourdainAl07}.}. Moreover, texts are freely edited and redistributed, and the encyclopedia has been developed with free and open source software. The entire editorial process, from the writing articles to the macrostructure organization, is collectively managed. Finally, the wikipedians have to respect elementary good manners. So, even if the Wikipedia editorial process totally differs from the traditional encyclopedia one, the goals of encyclopedic relevance and objectivity are in fact very close \cite{Endrezzi07,Giles05}.

Several formal and informal ways to regulate and control the encyclopedia have progressively been introduced by the wikipedian community in order to obey and to make users obey the \textit{pillars}. The common wikipedian philosophy makes it possible to gather together a large population of users writing about an unlimited number of themes or domains, to share their incomplete knowledge, to represent the various ways of thinking, and to delete errors thanks to successive users rectifications \cite{ViegasAl07,BryantForteBruckman05}. However, this philosophy also generates disputes and conflicts linked to inevitable disagreements between contributors. What processes does the wikipedian community use to resolve the conflicts, and what roles do the wikipedians choose in this process?

In this article, by analyzing the contributors behavior in places where conflicts are resolved, we provide elements to help answer these questions. The users behavior is observed both in the articles that are tagged as being in particular accordance (\textit{good} or \textit{featured articles}), or conversely not in accordance, with the main guidelines of Wikipedia (\textit{relevance dispute articles}, \textit{NPOV dispute articles}\dots{}), and in pages specifically dedicated to serious personal conflicts, the \textit{Arbitration Committee} \cite{ZlaticAl06,StviliaAl05}. As a result, we present the following contributions:

First, we make a users typology according to parameters that bring to light their involvement in conflicts and their publishing and management activity in the encyclopedia. In particular, we establish relationships between the number of appearances before the Arbitration Committee, the initiation of a request to the Arbitration Committee, and the numbers of contribution to articles and talk pages of Wikipedia. We show that major contributors are often involved in arbitration, and mostly as the initiating party. 

Then, we analyse the distribution of those types of users among the contributors to articles that do not respect a neutral point of view, given that it is one of the most important principles of Wikipedia. We find that all the major contributors who take their conflict before the Arbitration Committee are also contributors to NPOV articles, against only one half for the minor contributors.

Finally, by analysing the distribution of those wikipedians involved in serious disputes, among the contributors to tagged articles, we find that major contributors who are often involved in arbitration, are much more frequently contributing to protected articles (subject to disputes or vandalism), than to featured articles.

\section{Related work}

A  number of  authors study  conflicts in  Wikipedia in  relation with
coordination  and  cooperation   underlying  collaborative  work.  For
instance, \cite{KitturAl07} develop  quantitative  measures  of the  costs
involved  by   collaborative  work,  using  the   concepts  of  direct
(i.e.  writing   article)  and  indirect  work   (i.e.  discussion  or
anti-vandalism). At the article level, the history of the revisions is
often used to model and identify conflict or coordination periods 
\cite{KitturAl07,ViegasAl04}. The aim of the present study is rather
to analyse the behavior of wikipedians, who are involved in conflicts,
faced with the main tools wikipedians use to resolve conflicts.

Studies  of   conflict  management  and  social   control  in  virtual
communities  show that  such  social  systems have  the  same kind  of
problems as real social  systems. In particular, \cite{KollockAl96} 
show  that  the  \textit{social  dilemma}  between  individual  and
collective interest in  the problem of cooperation remains,  even if it
takes other  forms.  Furthermore, \cite{DuVal99} observes  that methods
using  both mediation  and  arbitration better  manage conflicts  than
power strategies of social control, as  it does in the real world. 
Indeed, the way a community manages its conflicts reveals its governance
mode \cite{AurayAl07,KitturAl07,ViegasAl04}. In  the French  Wikipedia,
mediation takes place in talk pages of  articles which have a template 
message at the top  of the  page, and arbitration  takes place in  the 
Arbitration Commitee pages. 

In fact, template messages at the  top of article pages are strongly
linked to the official guidelines of  Wikipedia. Indeed, these 
principles play an important role in the management and resolution of 
conflicts. \cite{ViegasAl07} analysed the content of the article talk 
pages, and found  that 7.9\% of the activity in those pages consists 
in references to Wikipedia official guidelines.

The  behavior  of wikipedians  has  been  studied  either from  their
motivations  point of  view \cite{Kuznetsov06}, either considering the 
type \cite{OrtegaAl07} or the evolution of their participation 
\cite{BryantForteBruckman05}. Our analysis of the behavior of wikipedians 
is based on quantitative data as well as in \cite{OrtegaAl07}, but is
restricted to those wikipedians  who are involved in conflicts.

\section{Corpus}

Wikipedia is a generic term for the free multilingual and collaborative online encyclopedia\footnote{Available at \url{http://www.wikipedia.org/}.} as well as a reference to every instance of this encyclopedia. Each instance refers to a different country and/or language. The instance we are interested in for this article is the French version of Wikipedia\footnote{Available at \url{http://fr.wikipedia.org/}.}. The corpus we used was extracted from the Wikipedia backup of 2006/04/02: more than 600,000 pages including 370,000 article pages and 40,000 talk pages (according to Wikipedia's internal architecture, each article page can be linked to a talk page). A tool called Wiki2Tei\footnote{Open software available at \url{http://wiki2tei.sourceforge.net/} and freely distributed according to the terms of the BSD license (\url{http://www.opensource.org/licences/bsd-license.php}).} was then used in order to convert the wikitext syntax to a TEI-compliant XML syntax (TEI standing for \textit{Text Encoding Initiative}).

The articles of Wikipedia are written by voluntary contributors working with each other via a wiki. Since anyone can freely edit any article, many virtual places are provided to avoid or settle conflicts that may arise in the process. First of all, each article is linked to a discussion page where contributors can exchange and justify their assertions, and thus reach compromises according to Wikipedia's netiquette and neutrality policy. Furthermore, users can insert specific tags\footnote{Defined in Wikipedia as "a frame type in articles indicating a piece of information or a link" \url{http://en.wikipedia.org/wiki/Wikipedia:Template}.} on top of articles which do not respect Wikipedia's official guidelines (such as neutrality or relevance dispute) \cite{Kuznetsov06,ForteAl05} or, on the contrary, to reward an exemplary article (called \textit{featured} or \textit{good articles}\footnote{\url{http://en.wikipedia.org/wiki/Wikipedia:Good_articles}.}). Theses tags are used to highlight for the community the fact that some articles need improvement and thus can be used as points of reference for users. Finally, when disputes degenerate into personal conflicts and get out of hand, each user can register a complaint to the Arbitration Committee. The Arbitration Committee is a group composed of seven contributors to Wikipedia, elected by the rest of the community for six months. Deliberations and votes of the Arbitration Committee are public and usually tend to reach unanimity, which implies consensus, as it is the rule for the articles. The role of Arbitrators is not to express an opinion about the scientific rightness or the editorial policy of an article but to ensure that Wikipedia's official guidelines are respected: neutral point of view (NPOV), the need to cite general sources, netiquette (called \textit{wikilove} by the wikipedian French community), the respect of the law, etc. They have the right to impose sanctions on users such as temporary or definitive article probation (meaning that the user cannot contribute anymore to one or more articles) or, less often, general restriction (meaning that the user is literally banned from all Wikipedia).

Thus, there are three virtual places to manage a conflict, in order of seriousness: the discussion pages linked to an article, the discussion pages linked to an NPOV dispute article and the pages of the Arbitration Committee. We focus on the last two because they correspond to open conflicts.

The first corpus we collected is composed of about 1,000 articles that have (or have had) the NPOV tag. Each article is associated, when possible, to its discussion page (some articles are not linked to a discussion page because the discussion may have started after we extracted the corpus). About 1,600 contributors intervened in these pages. We automatically added semantic tags to this corpus in order to extract each contribution and its size, who wrote it and when -- which tells us which contributions were written during the conflict and which were not -- and, when possible, to whom it answers. However, it is impossible to know who wrote a contribution when users do not sign it, deliberately or not. This is the reason why between 2\% and 5\% of the contributions may have been improperly tagged.

The second corpus is composed of about 80 pages from the Arbitration Committee. These pages are relatively well formed and homogeneous, allowing us again to automatically tag them so as to clearly make their essential architecture stand out: the conflict description, who registered the complaint and when, the parties involved, if the complaint is admissible or not, and the verdict of the arbitrators. Furthermore, each user is associated to his messages, and each arbitrator to his contributions and, of course, his vote. Finally, the verdict is composed of at least one verdict proposal and a vote; there are as many counterproposals and votes as needed until the arbitrators are able to reach an agreement. Each proposal is clearly identified and associated to the right arbitrator and each vote is associated to its arbitrator and to the proposition it refers to.

\section{Typology of users in conflict}
\label{typology}

The  Arbitration  Committee  is  therefore  a  formal  place  for  the
resolution of conflicts.  Though rather rare -- only about one hundred
users among 31\,000 wikipedians  were implied in an arbitration within
a  5-year period  --,  arbitrations represent  an  important tool  for
Wikipedia   governance.   Indeed,   elected  arbitrators   can  impose
penalties    against   Wikipedia    users    who   transgressed    the
\textit{pillars}. For instance, penalty may consist in blocking a user
in order to  keep the user from writing within articles during a certain period of time.  It  therefore  gives  strong  means for  controlling
publication.

Among the hundred arbitrations which  took place from the beginning of
Wikipedia-France to  2006 april, some user names  appear  more often,
either as the \textit{initiating party}, or as the \textit{other involved party}.  Those
two topics, frequency of appearance  and role in the complaint, allow
us to draw  up an initial typology of users engaged  in a dispute.  We
first  distinguished  three  kind  of protagonists  depending  on  the
frequency of  their appearances:  \textit{very regular ones}  who have
between 3  et 14 appearances\footnote{14  is anyway a sort  of record,
then there are two of them  having 7, another having 4, the other ones
having 3 appearances}, \textit{regular ones} who have two appearances,
and \textit{occasional ones} who have only one appearance.  Concerning
their   role   in   the   complaint,   we   then   distinguished three categories,  the \textit{initiating party},  that is to say those who are most 
often the initiator of the complaints, the \textit{other involved party}, 
and finally those who appear in a more balanced way,  sometimes as 
initiating party and sometimes as other involved party.  We can see on  Table~\ref{tbl:1} that among the
wikipedians  who often  appear,  the \textit{very  regular ones},  are
the initiating party for  most  part, even though  \textit{occasional ones},
who appeared only once, are mainly other involved party.  We also note
that most of those who appeared twice took once the initiating party
position, and once the other involving party position.

\begin{table}[h]
\setlength{\tabcolsep}{2mm} \centering
\caption{Appearances before the Arbitration Committee}
\begin{tabular}{lcccc}
\hline
\multicolumn{1}{c}{Appearances} & Users & Initiating party & Other party & Both \\
\hline
3--14 (very regular ones)       & 10    & 50\%             & 30\%        & 20\% \\
2 (regular ones)                & 17    & 12\%             & 29\%        & 59\% \\
1 (occasional ones)             & 74    & 30\%             & 70\%        & 0\% \\
\hline
\end{tabular}
\label{tbl:1}%
\end{table}

We then added to that typology the way users contribute to Wikipedia. We considered the number of their contributions in editing articles,  either in article pages, or in the discussion pages, because it is mainly in this place that conflicts begin\footnote{We did not consider for instance contributions in the  \textit{bistrots}  of Wikipedia.}.   Concerning this point, we noted big differences between users.  We drew up four categories, the \textit{major contributors} whose number of contributions extends from about
12,000  to 40,000  during the studied period,  the \textit{Large contributors},   between   2,800   and   12,000   contributions,   the
\textit{middle contributors}   between   600   and  2,800,   and the
\textit{minor contributors},  between one and  600 contributions.  Finally, we considered the type of their contributions according to whether they contribute to article pages or discussion pages.  We therefore distinguished three categories according to whether they contribute more often to articles or to discussions, or to both of them in a balanced way.

\begin{table}[h]
\setlength{\tabcolsep}{2mm} \centering
\caption{The contributions of the protagonists before the Arbitration Committee}
\begin{tabular}{lcccc}
\hline
\multicolumn{1}{c}{Contributions} & Users & Article orient. & Discussion orient. & Both \\
\hline 
12,000--40,000 (Major contrib.)   &  7    & 100\%           & 0\%                & 0\% \\
2,800--12 000 (Large contrib.)    & 23    & 96\%            & 0\%                & 4\% \\
600--2,800 (Middle contrib.)      & 31    & 81\%            & 0\%                & 19\% \\
1--600 (Minor contrib.)           & 40    &  70\%           & 5\%                & 25\% \\
\hline
\end{tabular}
\label{tbl:2}%
\end{table}

Table~\ref{tbl:2} shows that users who get involved in disputes in
Wikipedia contribute  more to  articles  than  to the  associated
talk pages, despite their conflicts.  Nevertheless, it also shows that the less they contribute to articles, the more they have a tendency to discuss.

Comparing the number of contributions and the frequency of appearances
(Table~\ref{tbl:3}),  we realize that parties of the Arbitration Committee who are \textit{very regular} are for the most part  \textit{big
contributors}, while   \textit{occasional} ones  are more often
\textit{small contributors}.

\begin{table}[h]
\setlength{\tabcolsep}{2mm} \centering
\caption{Categories of contributors in complaints}
\begin{tabular}{lccccc}
\hline
\multicolumn{1}{c}{Appearances} & Contributors & Major & Large & Middle & Minor \\
\hline
3--14 (very regular ones)       & 10           & 20\%  & 50\%  & 30\%   & 0\% \\
2 (regular ones)                & 17           & 13\%  & 29\%  &  29\%  & 29\% \\
1 (occasional ones)             & 74           & 4\%   & 18\%  &  31\%  & 47\% \\
\hline
\end{tabular}
\label{tbl:3}%
\end{table}

Comparing the  number of contributions  and the role in  the complaint
(Table~\ref{tbl:4}),  we note that  the \textit{big  contributors} are
more   often  the   initiating  party   and  that   the  \textit{small
contributors} are more often the  other involved party. Indeed we note
an increase of  the proportion of \textit{other involved  party} and a decrease
of the proportion  of \textit{initiating party} as the  number of contributions
decreases.   Part of  protagonists  who are  sometimes the  initiating
party  and sometimes  the other  involved party  is marginal  for each
category of contribution size.

\begin{table}[h]
\setlength{\tabcolsep}{2mm} \centering
\caption{Role in the complaint by size of contribution}
\begin{tabular}{lcccc}
\hline
\multicolumn{1}{c}{Contributions} & Users & Initiating party & Other party & Both \\
\hline
12,000--40,000 (Major contrib.)   & 7     & 57\%             & 29\%        & 14\% \\
2,800--12,000 (Large contrib.)    & 23    & 39\%             & 44\%        & 17\% \\
600--2,800 (Middle contrib.)      & 31    & 32\%             & 58\%        & 10\% \\
1--600 (Minor contrib.)           & 40    & 15\%             & 75\%        & 10\% \\
\hline
\end{tabular}
\label{tbl:4}%
\end{table}

The analysis  of these  tables evokes  that the  big contributors
assimilated  the  \textit{pillars}  of Wikipedia,  and  really care  about
enforcing  them \cite{AnthonyAl05,ForteAl05}.  Indeed,  the emerging  trend is  that the  more they
contribute to articles, the more they carry out publication control at
the same  time.  They  exercise this control  in the framework  of the
Arbitration  Committee  through their  role as  initiating
party.  They exercise  this  control mainly  over \textit{middle}  and
\textit{small} contributors.

In the following  section, we study whether  we can complete this
typology of  contributors before the Arbitration Committee  with the types
of  article they contribute  to, involving  the \textit{pillars}  of Wikipedia.
Indeed, we saw that users  put different tags within articles in order
to warn other users about breaches  of the rules of Wikipedia. We used
those  tags to  categorize articles  as \textit{featured
articles}, \textit{NPOV dispute articles}, \textit{relevance dispute articles}, and
\textit{protected articles}.

\section{Users in conflict and \textit{pillars} of Wikipedia}

The NPOV dispute tag is the first tangible evidence of a disagreement between wikipedians. Thus we studied characteristics of contributors who participated in articles with the NPOV tag, and particularly the ones who are also parties of arbitration by the Arbitration Committee. This analysis reveals several behavior trends. In Table \ref{tbl:5}, we study the behavior of the contributors, shared out in categories following the number of their contributions. We compare contributors in articles with a NPOV tag to all the contributors in Wikipedia. The second column indicates for each section the number of contributors in NPOV articles. The third column shows the number of appearances before the Arbitration Committee for the contributors in NPOV articles in comparison with all the protagonists before the Arbitration Committee, for each category (see Table \ref{tbl:2}). In Table \ref{tbl:6}, we study the behavior of the contributors who appear before the Arbitration Committee, considering on the one hand the appearance frequency, and on the other hand their role in the complaint. The second column indicates, for each category of frequency and of role, the number of contributors in Wikipedia who appear before the Arbitration Committee. The third column indicates for each category the number of contributors in NPOV articles who appear before the Arbitration Committee, and the proportion of these contributors to all the contributors of the same category who appear before the Arbitration Committee. Table \ref{tbl:5} shows that 77\% of the protagonists before the Arbitration Committee appear among the 1600 contributors participating to at least one article with the NPOV tag. It suggests that a lot of conflicts arise from an objectivity controversy.

\begin{table}[h]
\setlength{\tabcolsep}{2mm} 
\centering
\caption{Protagonists who appear before the Arbitration Committee (AC) among the contributors in NPOV articles, by contributions size}
\begin{tabular}{lcc}
\hline
\multicolumn{1}{c}{Contributors categories}  & \multicolumn{1}{c}{\# NPOV contributors}  & \multicolumn{1}{c}{NPOV contributors before the AC} \\
\hline
Major contributors                           & 30                                        & 7 (100\% of 7) \\
Large contributors                           & 151                                       & 21 (91\% of 23) \\
Middle contributors                          & 335                                       & 27 (84\% of 31) \\
Minor contributors                           & 1121                                      & 23 (57\% of 40) \\
\hline
Total                                        & 1637                                      & 78 (77\% of 101)\\
\hline
\end{tabular}
\label{tbl:5}
\end{table}

\begin{table}[h]
\setlength{\tabcolsep}{2mm}
\centering
\caption{Protagonists who appear before the Arbitration Committee (AC) among the contributors in NPOV articles, by appearances type}
\begin{tabular}{lcc}
\hline
\multicolumn{1}{c}{Protagonists categories} & \multicolumn{1}{c}{Before the AC} & \multicolumn{1}{c}{In NPOV articles} \\
\hline
Very regular                                & 10                                & 10 (100\%)  \\
Regular                                     & 17                                & 12 (70\%) \\
Occasional                                  & 74                                & 56 (76\%)  \\
\hline
Initiating party                            & 29                                & 26 (90\%)  \\
Other party                                 & 60                                & 44 (73\%) \\
Both                                        & 12                                & 8 (67\%)  \\
\hline
\end{tabular}
\label{tbl:6}
\end{table}

We also notice (Table \ref{tbl:5}) a very marked presence of the protagonists who appear before the Arbitration Committee among the most verbose contributors of our sample. We also note (Table \ref{tbl:6}) that the \textit{very regular protagonists} before the Arbitration Committee and the initiating parties contribute more in NPOV pages than \textit{regular} and \textit{occasional protagonists}, or than other involved parties. The \textit{very regular protagonists} and initiating parties are particularly present in NPOV discussions.

In order to study further the behavior of the contributors in conflict, we now consider their participation in other articles with a particular tag, indicating either a breach of relevance or objectivity principles, or a particular ageement with the official guidelines of Wikipedia. These tags are the neutral point of view (NPOV) dispute tag, the relevance dispute tag and the protected article tag, that takes place when the controversy degenerates into conflict in order to prevent the article from being modified, and the featured article tag, that indicates its particular quality, according to the \textit{pillars}.

\begin{figure}[!ht]
\begin{center}
\caption{Contributors in protected articles and protagonists}
\includegraphics[width=10cm]{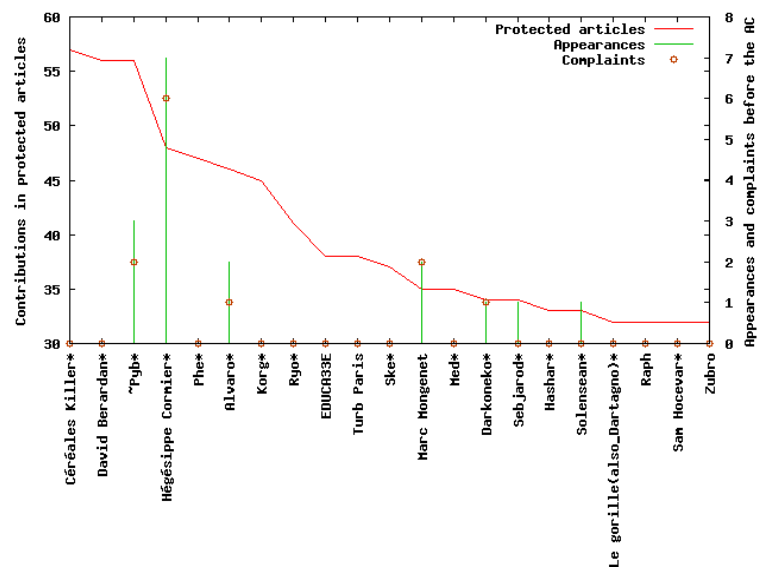}
 \label{protected}
\end{center}
\end{figure}

\begin{figure}[!ht]
\begin{center}
\caption{Contributors in featured articles and protagonists}
\includegraphics[width=10cm]{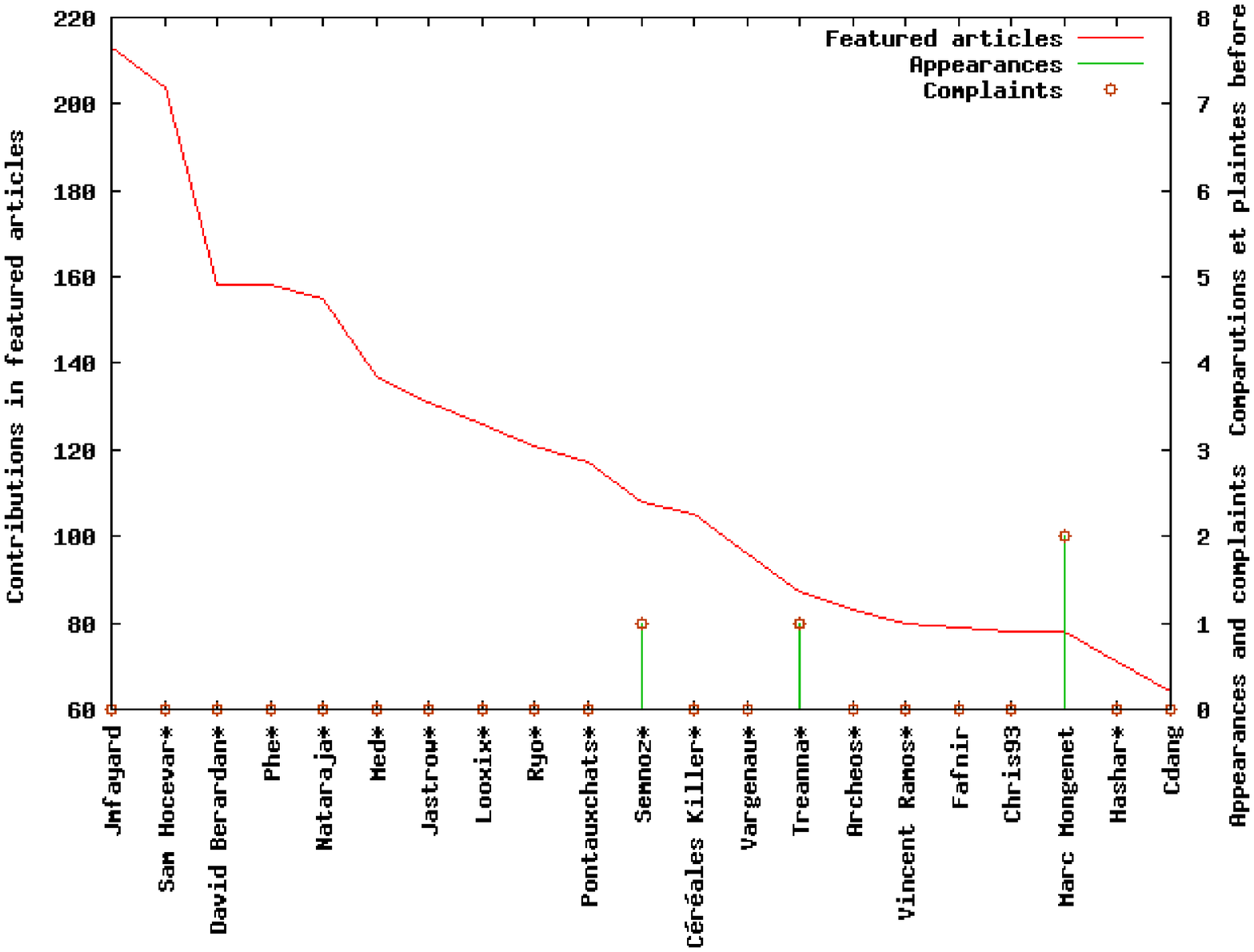}
 \label{featured}
\end{center}
\end{figure}

\begin{figure}[!ht]
\begin{center}
\caption{Contributors in NPOV articles and protagonists}
\includegraphics[width=10cm]{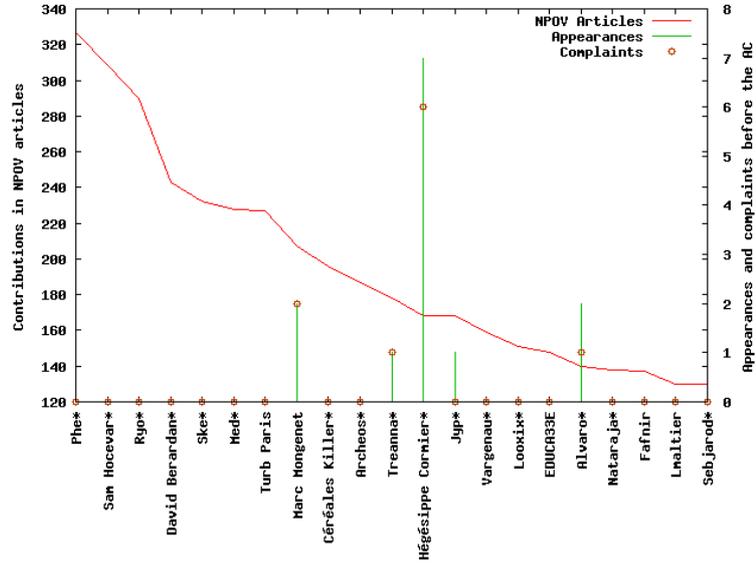}
 \label{NPOV}
\end{center}
\end{figure}

\begin{figure}[!ht]
\begin{center}
\caption{Contributors in non-relevant articles and protagonists}
\includegraphics[width=10cm]{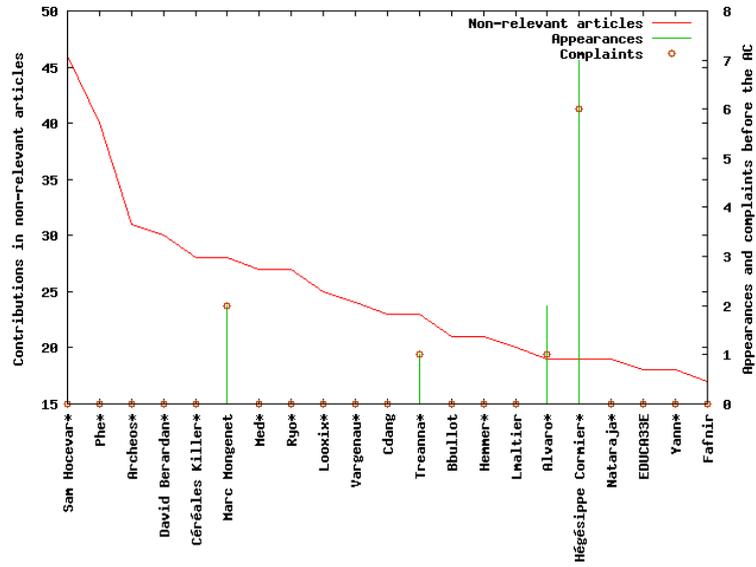}
\label{relevance}
\end{center}
\end{figure}

In Figures \ref{protected}, \ref{featured}, \ref{NPOV} and \ref{relevance}, the sample comprises only contributors in NPOV articles, who sometimes also contribute in articles with another tag. The curves in these figures  present in descending order the number of contributions for the 20 most verbose contributors, respectively in protected articles, in featured articles, in NPOV articles and in non-relevant articles. For each contributor, the number of his appearances before the Arbitration Committee (vertical line) and the number of his complaints (small circle) are also indicated, corresponding to the right scale.

We observe several interesting differences in these figures. In particular, among the 20 most verbose contributors in protected articles (Figure \ref{protected}), 7 are protagonists before the Arbitration Committee, namely 35\% of the major contributors on these articles. Furthermore, their behavior before the Arbitration Committee is disparate: some of them initiate the procedure and the others are other involved parties, some are very regular or regular protagonists and the others are occasional ones. On the other hand, Figure \ref{featured} shows that, among the most verbose contributors in featured articles, only 3 appeared before the Arbitration Committee, all of them as initiating parties. Nonetheless their apparent aggressiveness must be put into perspective: as none of these protagonists is a regular one, the complaints are few.

The behavior of the major contributors in NPOV and relevance dispute articles is between these two trends. Among the 20 most prolific contributors in NPOV articles indeed (Figure \ref{NPOV}), 25\% appeared before the Arbitration Committee. And 4 of the 20 major contributors in non-relevant articles, ie 20\%, also appeared in arbitrations (Figure \ref{relevance}).

In all these figures, the wikipedians with a particular status\footnote{Some particular status exists in the wikipedian community, e.g. administrator, steward, arbitrator, bureaucrat\dots{} Such a status is conferred by the community to a contributor through an election process. This status grants him/her extended rights in prospect of managing the encyclopedia.} are starred (*). It is interesting that most of the major contributors in the considered articles have also a particular status.

This observation confirms the previously mentioned correlation between a strong involvement of the contributors in the Wikipedia project, denoted both by the number of contributions and by the particular status \cite{AnthonyAl05,ForteAl05}, and their intervention where and when the official guidelines need to be protected.

\section{Conclusion}

The Wikipedia encyclopedia is mainly based on collaborative work. This
official guideline  yields   to  cooperation  patterns,  including
discussions  and information sharing  in order  to realize  the common
goal.    But   such   an   extended   collaboration   also   engenders
conflicts.  Disagreements  which   degenerate  into  serious  personal
disputes, with possible insults or systematic reverts, are finally not
so  frequent.   They only  involved  one  hundred  users among  30,000
wikipedians  over a  period of  five years.  Official  guidelines, the
Wikipedia \textit{pillars}, are clear, and there are not many
of them.  They constitute strong bases for  conflict resolution. Tools
and procedures  have been developed step  by step in  order to enforce
those principles.

We studied conflict evolution through the behavior of users who appear
before the  Arbitration Committee, and through  their contributions to
those  articles that  are tagged  such as  \textit{featured articles},
\textit{NPOV    articles},    \textit{non-relevant   articles},    and
\textit{protected articles}. As  expected, users appearing before the 
Arbitration Committee  are more numerous on articles subject  to a
NPOV or  relevance controversy, and  much more on  protected articles,
than on featured articles.

The presence of involved  parties before an Arbitration Committee has
different meanings depending on whether one is the initiating party or
the other involved  party. We note that major  and large contributors,
also often  involved as Wikipedia  administrators, do most  of the
job of publication control.  They are more often the ones who initiate
arbitrations,  and  moreover the  ones  who  contribute  the most  to
featured  articles.  Tables~\ref{tbl:2},  \ref{tbl:3},  \ref{tbl:4} of
Section  \ref{typology}   clearly  show  the   evolution  of  the   relative  sizes
respectively  between initiating parties  and other  involved parties,
between  contribution  to articles  and  contribution to  discussions,
between  regular and  occasional involved  parties  before Arbitration
Committee, according to the size of contributions.

As a result, we may say  that conflicts in Wikipedia are resolved both
by  means  of a  strong  commitment  to  clear official guidelines,
through specific places devoted to managing them, and by interventions
of some attentive users.

\subsubsection*{Acknowledgements}

The research reported here was supported by a grant from the French National Research Agency (ANR), within the framework of the Autograph project ANR-05-RNRT-03002 (S0604108 W).

%


\begin{thebibliography}{5}


\bibitem{AnthonyAl05} Anthony, D., Smith, S., Williamson, T.: Explaining 
Quality in Internet Collective Goods: Zealots and Good Samaritans in the 
Case of Wikipedia. Dartmouth College, Hanover (2005)

\bibitem{AurayAl07} Auray, N., Poudat, C., Pons, P.: Democratizing 
scientific vulgarization. The balance between cooperation and conflict 
in French Wikipedia. Observatorio 3, 185--199 (2007)

\bibitem{BryantForteBruckman05} Bryant, S.L., Forte, A., Bruckman, A.: 
Becoming Wikipedian: transformation of participation in a collaborative 
online encyclopedia. In: ACM SIGGROUP Conference on Supporting Group 
Work, pp. 1--10. ACM Press, New York (2005)

\bibitem{DuVal99} DuVal Smith, A.: Problems of Conflict Management in
Virtual Communities. In: Smith, M., Kollock, P. (eds.) Communities in 
Cyberspace, pp. 134--166. Routledge, London (1999)

\bibitem{Endrezzi07} Endrezzi L.: La communauté comme auteur et éditeur: 
l'exemple de Wikipédia. In: Jounée d'étude des URFIST "Évaluation et 
validation de l'information sur Internet" (2007)

\bibitem{ForteAl05} Forte, A., Bruckman, A.: Why Do People Write for 
Wikipedia? Incentives to Contribute to Open-Content Publishing. In: GROUP 05 
Workshop: Sustaining Community: The Role and Design of Incentive Mechanisms 
in Online Systems (2005)

\bibitem{Giles05} Giles, J.: Internet encyclopaedias go head to head. 
Nature 438(7070), 900--901 (2005)

\bibitem{GourdainAl07} Gourdain, P., O'Kelly, F., Roman-Amat, B., 
Soulas, D., von Droste zu Hülshoff, T.: La Révolution Wikipédia. Les 
encyclopédies vont-elles mourir? Mille et Une Nuits, Paris (2007)

\bibitem{KitturAl07} Kittur, A., Suh, B., Pendleton, B.A., Chi, E.H.: He 
says, She Says: Conflict and Coordination in Wikipedia. In: SIGCHI 
Conference on Human Factors in Computing Systems, pp. 453--462, ACM 
Press, New York (2007)

\bibitem{KollockAl96} Kollock, P., Smith, M.,: Managing  the  virtual  commons:
cooperation and conflict in Computer communities. In: Computer-Mediated
Communication:  Linguistic,  Social  and Cross-Cultural  Perspectives,
Susan Herring (ed.), pp. 109--128. John Benjamins, Amsterdam (1996)

\bibitem{Kuznetsov06} Kuznetsov, S.: Motivations of contributors to 
Wikipedia. ACM SIGCAS Computers and Society 36(2), 1--7 (2006)

\bibitem{OrtegaAl07} Ortega, F., Gonzalez-Barahona, J.M.: Quantitative 
Analysis of the Wikipedia Community of Users. In: WikiSym'07, pp. 75--86,
Montreal, Canada (2007)

\bibitem{StviliaAl05} Stvilia, B., Twidale, M., Gasser, L., Smith, L.: 
Information Quality Discussions in Wikipedia. Technical Report, 
University of Illinois at Urbana-Champaign (2005)

\bibitem{ViegasAl04} Viégas, F.B., Wattenberg, M., Dave, K.: Studying 
Cooperation and Conflict between Authors with history flow 
Visualizations. In: SIGCHI Conference on Human Factors in Computing 
Systems, pp. 575--582. ACM Press, New York (2004)

\bibitem{ViegasAl07} Viégas, F.B., Wattenberg, M., Kriss, J., Van Ham, 
F.: Talk Before You Type: Coordination in Wikipedia. In: 40th Hawaii 
International Conference on System Sciences (2007)

\bibitem{ZlaticAl06} Zlatic, V., Bozicevic, M., Stefancic, H., Domazet, 
M.: Wikipedias: Collaborative web-based encyclopedias as complex 
networks. Physical Review E, 74(1) 6--11 (2006)


\end{thebibliography}
\end{document}